\documentclass[
 aip,
 amsmath,amssymb,
 reprint,%
]{revtex4-1}

\usepackage{graphicx}
\usepackage{dcolumn}
\usepackage{bm}
\usepackage{float}

\usepackage[utf8]{inputenc}
\usepackage[T1]{fontenc}
\usepackage{mathptmx}
\usepackage{etoolbox}
\usepackage{newtxtext}
\usepackage[varvw]{newtxmath}
\usepackage{color,soul}
\usepackage[colorlinks=true,linkcolor=blue, citecolor= blue, urlcolor = blue]{hyperref}
\usepackage{cancel}

\usepackage{soul,xcolor}
\setstcolor{red}

\makeatletter
\def\@email#1#2{%
 \endgroup
 \patchcmd{\titleblock@produce}
  {\frontmatter@RRAPformat}
  {\frontmatter@RRAPformat{\produce@RRAP{*#1\href{mailto:#2}{#2}}}\frontmatter@RRAPformat}
  {}{}
}%
\begin{document}

\preprint{AIP/123-QED}

\title[]{Observation of astrophysically-relevant superadiabaticity in a plasma confined by a dipole magnet}
\affiliation{Department of Physics, Indian Institute of Technology Kanpur, Kanpur, Uttar Pradesh - 208016, India}

\author{Ayesha Nanda}
\altaffiliation[] {ayesha@iitk.ac.in}
\author{Sudeep Bhattacharjee}
\altaffiliation[] {sudeepb@iitk.ac.in}
\date{\today}
\begin{abstract}

The polytropic index of electrons in a magnetized plasma is experimentally investigated in the presence of heating and anisotropic work done, incorporating the effective dimensionality arising from temperature anisotropy. The study is performed in a plasma confined by a permanent cylindrical dipole magnet using a compact device. The measurements clearly demonstrate localized regions of superadiabatic electrons due to particle acceleration and energization through non-resonant heating processes. The realization of superadiabaticity is universal where heating dominates the work done and has broader implications for energy exchange processes in  space and astrophysically-relevant plasmas.

\end{abstract}

\maketitle
\begin{figure}[h]
\includegraphics[width=8.5 cm, height=3cm]{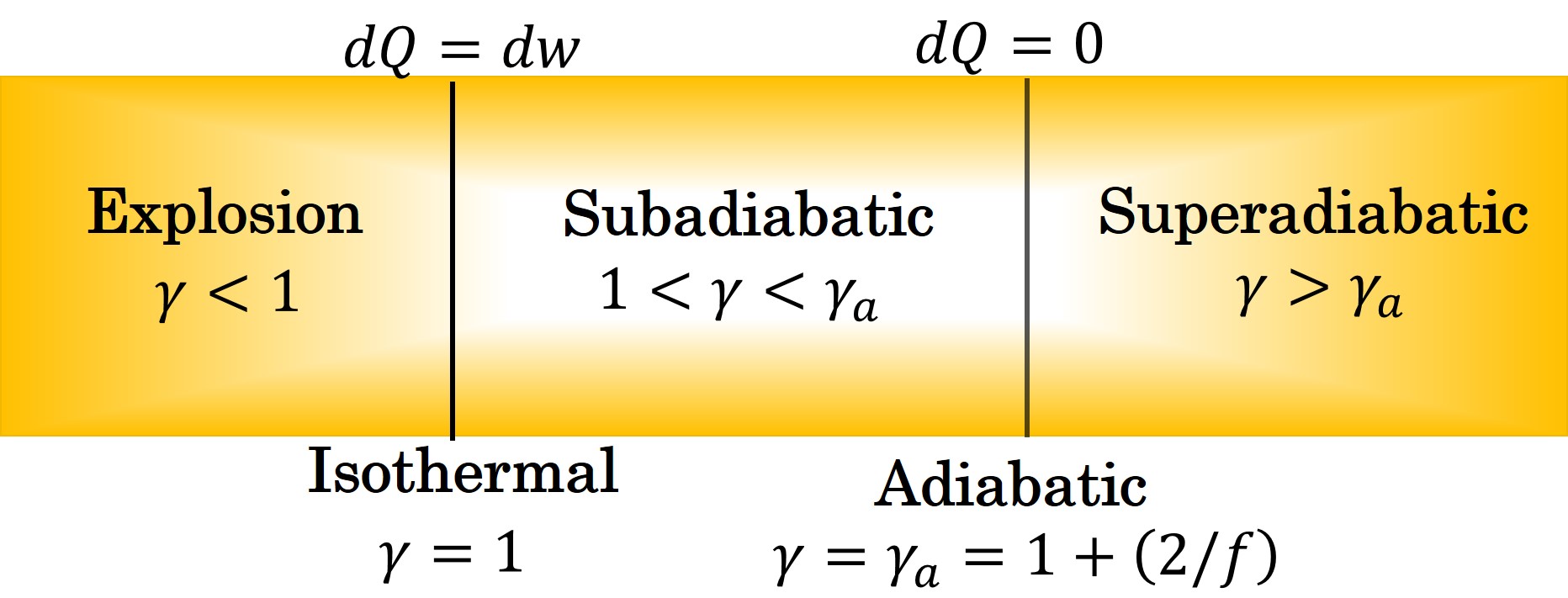}
\caption{\label{fig:pictorial}  Illustration of the polytropic index spectrum: $dQ$ is the heat applied and $dw$ is the work done.}
\end{figure}

Thermodynamic methods have been extensively applied in plasmas to investigate the exchange of heat, energy, and work \cite{frank2012plasma}. 
The polytropic index ($\gamma$) is an important parameter that characterizes compression or expansion and the associated heat transfer processes \cite{dayeh2022polytropic}, and has been the subject of extensive investigation in space and laboratory plasmas \cite{shaikh2023evidence,stasiewicz2005ion,little2016,takahashi2018}. The polytropic index varies widely, for example, astrophysical plasmas can have $\gamma < 1$ \cite{dialynas2018energetic}, or $\gamma$ close to the adiabatic index $\gamma_a$ ($= 1 + 2/f$, where $f$ is the kinetic degrees of freedom) \cite{wang2015evidence}, $\gamma$ can also be within the subadiabatic range ($1 < \gamma < \gamma_a$) \cite{prasad2018polytropic} or even be superadiabatic ($ \gamma > \gamma_a$) \cite{shaikh2023evidence}. In contrast, laboratory experiments typically yield polytropic indices that range from isothermal ($\gamma=1$) to adiabatic index, thus $\gamma \in [1, \gamma_a$] \cite{takahashi2020,zhang2016polytropic,kim2018thermodynamics, takahashi2019helicon}. The pictorial illustration of the polytropic index spectrum as per current understanding is shown in Fig. \ref{fig:pictorial}.

In all the aforementioned works, the polytropic index has either been studied in space or in laboratory magnetized plasmas other than dipole geometry. In the dipole plasma devices in laboratory, the main experimental findings include high beta plasma confinement \cite{saitoh2011,garnier2006production} and inward diffusion \cite{boxer2010} in Levitated Dipole Experiment (LDX) at MIT; chaotic particle transport \cite{warren1995observation}, collisionless radial transport \cite{mauel1992experiment} and generation of artificial radiation belt \cite{mauel1997laboratory} in Collisionless Terrella Experiment (CTX) at Columbia University; vortex formation \cite{yoshida2010} in Ring Trap-1 (RT-1) at the University of Tokyo; excitation of whistler \cite{huang2019simulations} and chorus \cite{huang2018excitation} waves in Dipole Research EXperiment (DREX) in China; and diffusion induced transport \cite{baitha2020steady}, optical emissivity \cite{bhattacharjee2021characterizing}, anisotropic electrical conductivity \cite{nanda2022}, current density profiles \cite{nanda2023}, energy distributions \cite{hunjan2023measurement} and fluctuations \cite{hunjan2024characteristics} in the table top device in our laboratory \cite{bhattacharjee2022physics}. Despite this extensive research, to the best of our knowledge, superadiabaticity \cite{shaikh2023evidence} has neither been observed in the laboratory nor investigated in detail in space.

\begin{figure*}[]
\centering
\includegraphics[width= 14.5 cm, height=10.3cm]
{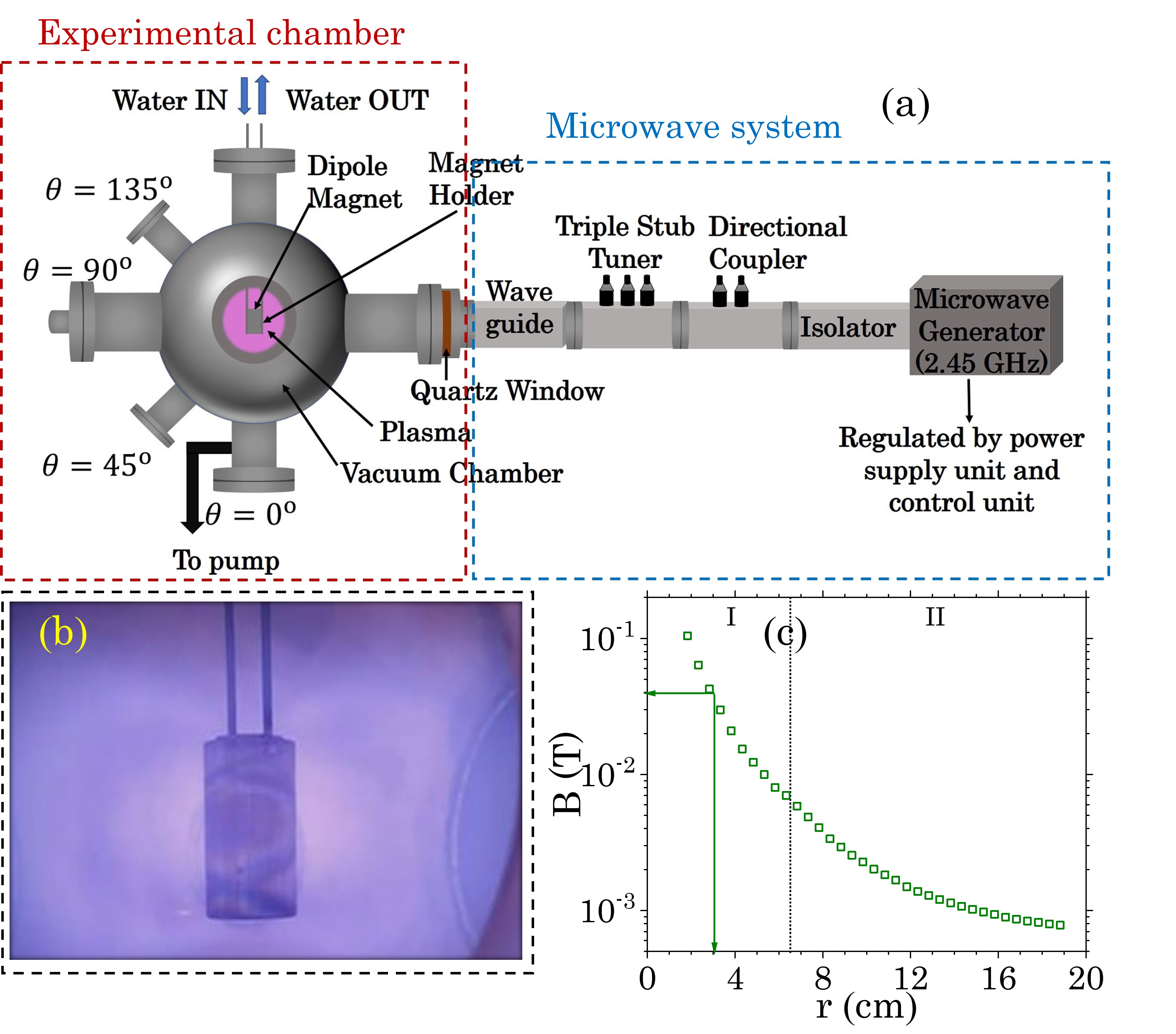}
\caption{\label{Fig_expt} (a) Schematic of the experimental set-up consisting of the experimental chamber and microwave system. (b) Digital picture of argon plasma confined by a permanent dipole magnet inside the experimental chamber, composed of local thermodynamic systems for the present study. (c) Measured magnetic field strength with radial distance at $\theta =90^\circ$ using a Lakeshore 421 gaussmeter with least count of 1
Gauss. Solid lines correspond to the radial location and magnetic field, where plasma frequency is equal to the electron cyclotron frequency.} 
\end{figure*}

In this Letter, we report for the first time the experimental observation of superadiabatic electrons in a magnetized laboratory plasma in the presence of  heating and anisotropic work done along various directions, incorporating effective dimensionality ($f^*$) \cite{livadiotis2021relationship}. Unique to the present study is the local obtainment of superadiabatic polytropic indices, driven by non-resonant heating processes as discussed later.

The experiment has been performed in a compact dipole device, as illustrated in Fig. \ref{Fig_expt}(a). The experimental set-up comprises a spherical vacuum chamber, evacuated to a base pressure of 1 $\mu$Torr through the combined action of a turbo molecular pump and a dry scroll pump. Argon is used as a test gas. Plasma generation is achieved through electron cyclotron resonance (ECR) heating using waves of 2.45 GHz generated by a magnetron-based microwave generator. The plasma is confined by a permanent cylindrical magnet (NdFeB) of length 4.1 cm and diameter 2.3 cm. The magnet is housed inside a water-cooled holder and is suspended from the top to the center of the experimental chamber (cf. Fig. \ref{Fig_expt}(b)). The magnetic field ($B$) profile of the magnet with $r$ at $\theta=90^\circ$ is shown in Fig. \ref{Fig_expt}(c). The vertical and horizontal solid lines correspond to the radial position (in cm) and magnetic field (in T) where $\omega_p = \omega_c$, which is attained at 3.1 cm at $\theta = 90^\circ$ (cf. Fig. \ref{Fig_expt}(a)); $\omega_p$ ($=N_ee^2/(m_e\epsilon_0)$) and $\omega_c$ ($=eB/m_e$) being the plasma and electron cyclotron frequencies, respectively. Here, $N_e$ is the measured electron density, $\epsilon_0$ is the permittivity of free space, $e$ and $m_e$ are the electronic charge and mass respectively. The plasma is magnetically confined until $\sim$ 20 cm from the center of the chamber. The confinement is notably stronger until a point ($r \sim 8$ cm), beyond which the space potential starts to decrease \cite{bhattacharjee2022physics,baitha2020steady}. Details of the polar profile of magnetic field intensity and ECR location can be found in Ref. \cite{baitha2019particle,bhattacharjee2022physics}.  In table \ref{Table}, $N_e$, perpendicular and parallel electron temperature ($T_{e\perp}$ and $T_{e\parallel}$), the characteristic magnetic field scale length $(|B/\nabla B |)$, electron Larmor radius ($r_L$ ($=\sqrt{m_ek_BT_{e\perp}}/eB$, $k_B$ being the Boltzmann constant)) and $\omega_p/\omega_c$ are tabulated at fixed radial distances selected from each region: I and II (separated by vertical dashed line in Fig. \ref{Fig_expt}(c)) for $\theta = 90^\circ$. The diagnostics used in this work are Langmuir probe with appropriate Bohm \cite{bhattacharjee2022physics} and geometric correction \cite{baitha2019particle} to the probe collection area due to the magnetized plasma, and the linear antenna for wave electric field measurements using micro-coax microwave grade RG316 (DC—3 GHz) cables \cite{nanda2023}. The measured electric field ($E_1^{'}$) is obtained from the average intensity $(1/2)\epsilon_0cE_1^{'2} = P/A$, where $c$ is the velocity of light, $P$ is the power detected by the antenna and $A$ ($=4.91\times10^{-6}$ m$^2$) is the area of the antenna. The measured $E_1^{'}$ is calibrated with a known electric field ($E_1$) inside a cylindrical waveguide with cross-section area $1.91 \times 10^{-2}$ m$^2$ subjected to varying input wave power (300$-$400 W) at 2.45 GHz frequency in vacuum. $E_1$ and $E_1^{'}$ are related as
\begin{equation}
    E_1 = (\log{E_1^{'}} + 6.51) \times 4.03 \times 10^2 \hspace{2mm} \text{V/m}.
\end{equation}
Using this relation, the antennas have been calibrated to obtain the exact microwave electric field inside the plasma.

The operating conditions for the present study are systematically controlled at the pressure of 0.4, 1.2 and 2.0 mTorr and input wave power of 300, 351 and 399 W. Measurements were taken at different polar angles $\theta = 0^\circ, 45^\circ, 90^\circ$ and $135^\circ$, and the results for 1.2 mTorr, 300 W at $\theta=90^\circ$ are presented. The construction of the diagnostics and operation details of the experimental
system can be found in the works of Nanda \textit{et al} \cite{nanda2023,nanda2022}.

\begin{table}[h]
\begin{center}
\caption{Electron density ($N_e$), perpendicular and parallel electron temperature ($T_{e\perp}$ and $T_{e\parallel}$), magnetic field scale length ($|B/\nabla B |$), Larmor radius ($r_L$) and the ratio of plasma frequency ($\omega_p$) to electron cyclotron frequency ($\omega_c$) at fixed radial distances from two regions: I (r = 3 cm) and II (r = 14 cm).}
\label{Table}
\begin{tabular}{*{7}{c}}
\toprule\\
\multicolumn{1}{ p{1 cm}}{\centering{r \\ (in cm)}}&\multicolumn{1}{p{1.5 cm}}{\centering {$N_e (\times 10^{16})$ \\ (in m$^{-3}$)}}&\multicolumn{1}{p{1 cm}}{\centering {$T_{e\perp}$ \\ (in eV)}}&\multicolumn{1}{p{1 cm}}{\centering {$T_{e\parallel}$ \\ (in eV)}}&\multicolumn{1}{p{1 cm}}{\centering {$|B/\nabla B |$\\ (in cm)}}&\multicolumn{1}{p{1 cm}}{\centering {$r_L$ \\ (in cm)}} &\multicolumn{1}{p{1 cm}}{\centering {$\omega_p / \omega_c$}} \\
\cline{1-7}\\
\multicolumn{1}{ p{1 cm}}{\centering{3 }}&\multicolumn{1}{p{1.5 cm}}{\centering {1.28}}&\multicolumn{1}{p{1 cm}}{\centering {4.79}}&\multicolumn{1}{p{1 cm}}{\centering {6.29}}&\multicolumn{1}{p{1 cm}}{\centering {1.26}}&\multicolumn{1}{p{1 cm}}{\centering {0.02}} &\multicolumn{1}{p{1 cm}}{\centering {0.86}} \\
\multicolumn{1}{ p{1cm}}{\centering{14} }&\multicolumn{1}{p{1.5 cm}}{\centering {1.72}}&\multicolumn{1}{p{1 cm}}{\centering {1.44}}&\multicolumn{1}{p{1 cm}}{\centering {1.63}}&\multicolumn{1}{p{1 cm}}{\centering {8.73}}&\multicolumn{1}{p{1 cm}}{\centering {0.38}} &\multicolumn{1}{p{1 cm}}{\centering {37.04}} \\
\botrule
\end{tabular}
\end{center}
\end{table}

The polytropic index in a plasma can be obtained from \cite{shaikh2023evidence,livadiotis2021relationship},
\begin{equation}
    \gamma_{\perp \text{or} \parallel} = 1 + (\gamma_a^*-1)\Big[1- \Big(\frac{dQ}{dw}\Big)_{\perp \text{or} \parallel}\Big],
    \label{eq:gamma}
\end{equation}
where $\perp$ and $\parallel$ denote the perpendicular and parallel direction to the magnetic ﬁeld ($B$) respectively, $dQ$ is the heat supplied to the system, $dw$ is the work done having positive (or negative) values signifying plasma expansion (or compression) and $\gamma_a^*$ ($=1 + (2/f^*)$) is the modified adiabatic index in an anisotropic plasma, $f^*$ being the effective dimensionality, which depends upon kinetic degrees of freedom $f$ and anisotropicity $\alpha$ ($=T_{e\perp} / T_{e\parallel}$), given by \cite{livadiotis2021relationship}, 
\begin{equation}
    f^* = \frac{1 + (f-1)^2\alpha^2}{1 + (f-1) \alpha^2},
    \label{eq:f^*}
\end{equation}
and holds good for $\alpha <2$ \cite{livadiotis2021relationship}. In our experiments, $\alpha$ lies between 0.7$-$1.7 \cite{nanda2022} and $f^*$ varies between 1.49$-$1.85 for $f=3$. Furthermore, the perpendicular and parallel work in the presence of $B$ are given by \cite{du2020energy,guo2017},
\begin{align}
     dw_\perp = p_\perp dV_\perp = -p_\perp V d(\ln B) \hspace{3 mm} \nonumber\\ \text{and} \hspace{3 mm} dw_\parallel =p_\parallel dV_\parallel = -p_\parallel V d \Big(\ln \frac{N_e}{B}\Big),
     \label{eq:work}
\end{align}
respectively, with $V$ being the  elementary flux tube volume. Substituting perpendicular and parallel plasma pressures $p_\perp (=N_ek_BT_{e\perp})$ and $p_\parallel (=N_ek_BT_{e\parallel})$, and $V$ ($=N/N_e$), $N$ being the total number of electrons in Eq. \ref{eq:work}, the ratio of net heat supplied to the work done (cf. Eq. \ref{eq:gamma}) can be derived as,
\begin{align}
    \Big(\frac{dQ}{dw}\Big)_\perp = \frac{N(dq_\perp-dl_\perp)}{dw_\perp} = -\frac{1}{k_B T_{e\perp}} \frac{(dq_\perp/dt-dl_\perp/dt)}{d(\ln B)/dt},
    \label{eq:dq_by_dw_perp}
\end{align}
and 
\begin{align}
    \Big(\frac{dQ}{dw}\Big)_\parallel = \frac{N(dq_\parallel-dl_\parallel)}{dw_\parallel}   = -\frac{1}{k_B T_{e\parallel}} \frac{(dq_\parallel/dt-dl_\parallel/dt)}{d(\ln (N_e/B))/dt},
    \label{eq:dq_by_dw_para}
\end{align}
where ($dq_{\perp \text{or} \parallel}/dt$) and ($dl_{\perp \text{or} \parallel}/dt$) are the heating and loss rates per electron respectively. For a plasma at steady-state confined in a static magnetic field,
\begin{align}
\label{eq:dq_perp_by_dt}
    \frac{dq_\perp}{dt} = \underbrace{\frac{e E_{1\perp}^2}{2m_e}\frac{\nu_{h\perp}}{\nu_{h\perp}^2 + (\omega - \omega_c)^2}}_{W_{M\perp}} + \underbrace{\frac{T_{e\perp}}{B} \Big( \vec{v}_E\cdot\vec{\nabla}B \Big )}_{W_{G\perp}},
\end{align}
\begin{align}
\label{eq:dq_para_by_dt}
    \frac{dq_\parallel}{dt} = \underbrace{\frac{e E_{1\parallel}^2}{2m_e}\frac{\nu_{h\parallel}}{\nu_{h\parallel}^2 + (\omega - \omega_c)^2}}_{W_{M\parallel}} + \underbrace{\Big(T_{e\parallel} + \frac{m_e}{e} v_{\parallel}^2 \Big) \vec{v}_E \cdot \Big(\hat{B} \cdot \vec{\nabla}\hat{B} \Big)}_{W_{C\parallel}},
\end{align}

\begin{align}
\label{eq:dl_perp_by_dt}
    \frac{dl_\perp}{dt} =  \underbrace{\nu_{z\perp}\epsilon_{z} + \nu_{x\perp }\epsilon_{x}}_{W_{in\perp }} + \underbrace{\nu_{h\perp}\frac{3m_e}{M}T_{e\perp}}_{W_{el\perp}}  + \underbrace{\nu_{c\perp} T_{e\perp}}_{W_{cx\perp }}  + \underbrace{\nu_{i\perp} (T_{e\perp} - T_{e \parallel})}_{W_{is\perp}},
\end{align}
\begin{align}
\label{eq:dl_para_by_dt}
    \frac{dl_\parallel}{dt} =  \underbrace{\nu_{z\parallel}\epsilon_{z} + \nu_{x\parallel }\epsilon_{x}}_{W_{in\parallel}} + \underbrace{\nu_{h\parallel}\frac{3m_e}{M}T_{e\parallel}}_{W_{el\parallel}}  + \underbrace{\nu_{c\parallel} T_{e\parallel}}_{W_{cx\parallel}}  + \underbrace{\nu_{i\parallel} (T_{e\parallel} - T_{e \perp})}_{W_{is\parallel}},
\end{align}
where $E_{1(\perp \text{or} \parallel)}$ is the wave electric field, $\nu_{h (\perp \text{or} \parallel)}$, $\nu_{z (\perp \text{or} \parallel)}$, $\nu_{x (\perp \text{or} \parallel)}$, $\nu_{c (\perp \text{or} \parallel)}$ and $\nu_{i (\perp \text{or} \parallel)}$ are the electron-neutral collision, ionization, excitation, charge exchange and isotropization frequencies respectively \cite{pandey2015particle,kawazura2015observation}, $\omega$ is the angular wave frequency, $v_{\parallel}$ is the fluid velocity obtained by solving the momentum equation along parallel direction \cite{nanda2022}, $M$ is the ionic mass, and $\epsilon_{z}$ and $\epsilon_{x}$ are the ionization and electronic excitation energies. The loss rate comprises losses due to inelastic \(W_{in(\perp \text{or} \parallel)}\) and elastic collisions \(W_{el(\perp \text{or} \parallel)}\) \cite{pandey2015particle}, charge exchange \(W_{cx(\perp \text{or} \parallel)}\) \cite{bhattacharjee2022physics} and isotropization \(W_{is(\perp \text{or} \parallel)}\) \cite{kawazura2015observation}. $W_{M(\perp \text{or} \parallel)}$ represents the rate of direct wave-induced heating dependent on its local electric field intensity; from henceforth referred to as wave-induced heating rate \cite{bhattacharjee2001plasma,bhattacharjee2007quasisteady}. Apart from the wave-induced heating, electrons undergoing grad$-B$ and curvature drifts along perpendicular and parallel directions respectively get accelerated by the electric field resulting in net energy gain, hereafter referred to as gradient ($W_{G\perp}$)  and curvature ($W_{C\parallel}$) heating rates \cite{tenbarge2024electron}. In $W_{G\perp}$ and $W_{C\parallel}$ (cf. Eq. (\ref{eq:dq_perp_by_dt}) and (\ref{eq:dq_para_by_dt})),   $\vec{v}_E$ is the $\vec{E}\times \vec{B}$ drift velocity arising from $B$ and the total electric field $\widetilde{\vec{E}} (\vec{r},t) = \vec{E_0}(\vec{r}) + \widetilde{\vec{E_1}}(\vec{r},t)$, where $\vec{E_0}(\vec{r})$ is the electrostatic field resulting from the gradient of the space potential and $\widetilde{\vec{E_1}}(\vec{r},t)$ is the wave electric field \cite{nanda2023}. In the present experiment, $\widetilde{\vec{E}} \cong \widetilde{\vec{E_1}}$ since $|\vec{E_0}|<<|\widetilde{\vec{E_1}}|$, and the drift velocity $\vec{v}_E$ is primarily governed by $\widetilde{\vec{E_1}}$. However, situation may arise where $\vec{E} \cong \vec{E_0}$ when $\widetilde{\vec{E_1}}$ is either absent or weak in comparison to $\vec{E_0}$. The derivatives in the denominator of the right hand side of Eq. \ref{eq:dq_by_dw_perp} and \ref{eq:dq_by_dw_para} can be obtained as,
\begin{align}
\label{eq:dB_by_dt}
    \frac{d(\ln B)}{dt} = \cancelto{0}{\frac{\partial (\ln{B})}{\partial t}} + v_\perp \nabla_\perp (\ln B),
\end{align}
\begin{align}
\label{eq:dNe/B_by_dt}
    \frac{d(\ln (N_e/B))}{dt} = \cancelto{0}{\frac{\partial (\ln{(N_e/B)})}{\partial t}} + v_\parallel \nabla_\parallel (\ln (N_e/B)),
\end{align}
where  $v_{\perp}$ is the fluid velocity along perpendicular direction \cite{nanda2022}.
Substituting Eq. (\ref{eq:dq_perp_by_dt}), (\ref{eq:dl_perp_by_dt}) and (\ref{eq:dB_by_dt}) in Eq. (\ref{eq:dq_by_dw_perp}), and Eq. (\ref{eq:dq_para_by_dt}), (\ref{eq:dl_para_by_dt}) and (\ref{eq:dNe/B_by_dt}) in Eq. (\ref{eq:dq_by_dw_para}), and then by employing Eq. (\ref{eq:dq_by_dw_perp}) and (\ref{eq:dq_by_dw_para}) in Eq. (\ref{eq:gamma}), the perpendicular and parallel polytropic indices can be derived as,
\begin{figure*}
\includegraphics[width=17 cm, height =12 cm]
{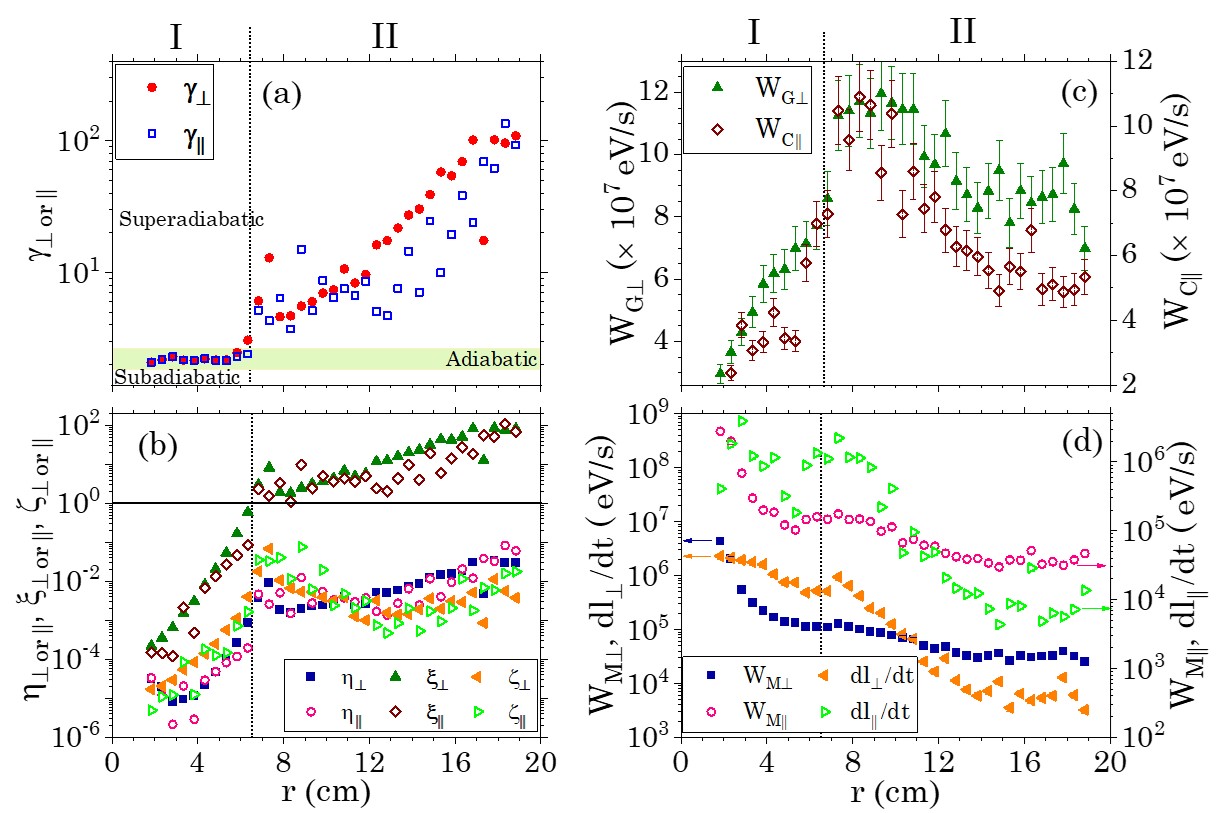}
\caption{\label{Fig_gamma}  
{Radial profiles of (a) polytropic index, $\gamma_{\perp \text{or} \parallel}$ with the green band denoting the adiabatic region, and (b) ratio of wave-induced heating to the work done, $\eta_{\perp \text{or} \parallel}$, ratio of gradient or curvature heating to the work done, $\xi_{\perp \text{or} \parallel}$ and ratio of energy loss to the work done, $\zeta_{\perp \text{or} \parallel}$. The error associated with the standard deviation is within 20$\%$ for \(\gamma_{\perp \text{or} \parallel}\), within 8$\%$ for \(\eta_{\perp \text{or} \parallel}\) and \(\xi_{\perp \text{or} \parallel}\), and within 11$\%$ for \(\zeta_{\perp \text{or} \parallel}\). Variation of (c) gradient, $W_{G\perp}$ (left y-axis) and curvature, $W_{C \parallel}$ (right y-axis) heating rates, and (d) perpendicular wave-induced heating rate $W_{M\perp}$ and energy loss rate $dl_\perp/dt$ (left y-axis) and parallel wave-induced heating rate $W_{M \parallel}$ and energy loss rate $dl_\parallel/dt$ (right y-axis)} with radial distance $r$ at 1.2 mTorr pressure and 300 W wave power in the equatorial plane ($\theta = 90^\circ$).
}
\end{figure*}
\begin{align}
   \gamma_\perp =  1 + (\gamma_a^*-1)\bigg [1+ \bigg(\underbrace{\frac{1}{k_B T_{e \perp}}\frac{W_{G \perp}}{v_\perp \nabla_\perp (\ln B)}}_{\xi_\perp}  \nonumber \\ + \underbrace{\frac{1}{k_B T_{e \perp}}\frac{W_{M \perp}}{v_\perp \nabla_\perp (\ln B)}}_{\eta_\perp} - \underbrace{\frac{1}{k_B T_{e \perp}}\frac{dl_\perp/dt}{v_\perp \nabla_\perp (\ln B)}}_{\zeta_\perp}\bigg) \bigg ],
   \label{Eq:gamma_perp}
\end{align}
\begin{align}
   \gamma_\parallel =  1 + (\gamma_a^*-1)\bigg [1+ \bigg(\underbrace{\frac{1}{k_B T_{e \parallel}}\frac{W_{C \parallel}}{v_\parallel \nabla_\parallel \ln (N_e/B)}}_{\xi_\parallel} \nonumber \\ + \underbrace{\frac{1}{k_B T_{e \parallel}}\frac{W_{M \parallel}}{v_\parallel \nabla_\parallel \ln (N_e/B)}}_{\eta_\parallel} - \underbrace{\frac{1}{k_B T_{e \parallel}}\frac{dl_\parallel/dt}{v_\parallel \nabla_\parallel \ln (N_e/B)}}_{\zeta_\parallel} \bigg) \bigg].
   \label{Eq:gamma_parallel}
\end{align}

Radial profile of $\gamma_{\perp \text{or} \parallel}$ is plotted in Fig.  \ref{Fig_gamma}(a) using Eq. \ref{Eq:gamma_perp} and \ref{Eq:gamma_parallel}. The polytropic index is \textit{nearly adiabatic} ($\gamma_{\perp \text{or} \parallel} \to \gamma_a^*$ where $\gamma_a^* =1+(2/f^*)= 1 + (2(1+2\alpha^2)/(1+4\alpha^2))$ for electrons with three kinetic degrees of freedom (cf. Eq. \ref{eq:f^*})) in region I and is \textit{superadiabatic} ($\gamma_{\perp \text{or} \parallel} > \gamma_a^*$) in region II. To understand the superadiabatic nature, radial profiles of the ratio of wave-induced heating to the work done ($\eta_{\perp \text{or} \parallel}$), gradient (or curvature) heating to the work done ($\xi_{\perp \text{or} \parallel}$) and energy losses to the work done ($\zeta_{\perp \text{or} \parallel}$) (cf. Eq. \ref{Eq:gamma_perp} and \ref{Eq:gamma_parallel}), are plotted in Fig. \ref{Fig_gamma}(b) at the same locations and operating conditions, as in Fig. \ref{Fig_gamma}(a). The work done along perpendicular and parallel directions is negative as $dV_{\perp \text{or} \parallel} <0$ (cf. Eq. \ref{eq:work}), and negative work implies plasma compression. Moreover, it is clearly observed that $\zeta_{\perp \text{or} \parallel}, \eta_{\perp \text{or} \parallel} < \xi_{\perp \text{or} \parallel}$ (cf. Fig. \ref{Fig_gamma}(b)), and hence gradient and curvature heating regulates the $\gamma_{\perp \text{or} \parallel}$ values. Near the magnet ($r <6.5$ cm), $\xi_{\perp \text{or} \parallel} << 1$ and $\gamma_{\perp \text{or} \parallel} \to \gamma_a^*$, and in regions at and beyond $r =6.5$ cm, $\xi_{\perp \text{or} \parallel} \geq 1$ resulting in $\gamma_{\perp \text{or} \parallel} > \gamma_a^*$ (cf. Fig. \ref{Fig_gamma}(a) and \ref{Fig_gamma}(b)). 

To assess the relevance of  heating and loss rates with the superadiabatic behavior of electrons, radial profiles of the rate of gradient ($W_{G \perp}$), curvature ($W_{C \parallel}$), wave-induced ($W_{M(\perp \text{or} \parallel)}$) heating and loss rates ($dl_{\perp \text{or} \parallel}/dt$) are shown in Fig. \ref{Fig_gamma}(c) and \ref{Fig_gamma}(d). In region II, the heating rates associated with gradient and curvature drifts ($W_{G \perp}$ and $W_{C \parallel}$) are relatively higher than the wave-induced heating ($W_{M(\perp \text{or} \parallel)}$) and loss rates ($dl_{\perp \text{or} \parallel}/dt$) by a factor of $\sim 10^2 - 10^4$ and $\sim 10^2 - 10^3$ in the perpendicular and parallel directions, respectively. This suggests that $W_{G \perp}$ and $W_{C \parallel}$ play a crucial role in regulating the net heat supplied to the plasma locally along their respective directions,  thereby regulating the values of $\gamma_{\perp \text{or} \parallel}$. From the observations, it can be concluded that the superadiabaticity is an outcome of the governance of non-resonant gradient and curvature heating.

In conclusion, this Letter investigates the polytropic index of electrons in a magnetized plasma, taking into account the effects of heating, anisotropic work done, and effective dimensionality arising from temperature anisotropy. The experiment demonstrates local regions with superadiabatic electrons for the first time in a laboratory magnetized plasma. Results indicate that the local superadiabatic electrons are due to heating through acceleration of electrons undergoing grad$-B$ and curvature drifts along perpendicular and parallel directions, respectively, which are intrinsic to magnetized plasma systems with gradients and curvatures in the magnetic field lines \cite{tenbarge2024electron}. Hence, the results are relevant for planetary magnetosphere with 3-D  magnetic bottle structures where particles are trapped \cite{nykyri2012origin,nykyri2019first}, undergoing gradient, curvature and $\vec{E} \times \vec{B}$ drift motion, as well as in solar wind physics \cite{sorathia2019solar} and heating mechanisms within solar wind transients \cite{burkholder2021magnetospheric}. Furthermore, comprehending these heating mechanisms is essential for predicting high-energy particle environments in space, which is crucial for safe space operations and preventing radiation hazards to astronauts \cite{nykyri2019first,burkholder2021magnetospheric}. 

The macroscopic study of the polytropic behavior enhances our understanding of the heating mechanisms in the energy exchange processes. The specific behavior of energy exchange in a superadiabatic system can vary widely based on the nature of the energy sources, associated particle drift velocities, the thermodynamical processes involved, and the plasma system's characteristics. Further, investigation of the thermodynamic properties will be taken up as future work.

\begin{acknowledgments}
The authors express their gratitude to the funding agency Council of Scientific and Industrial Research (CSIR), India, through Grant No. 03/1496/23/EMR-II. Ayesha Nanda gratefully acknowledges doctoral fellowship under the INSPIRE programme (IF180114) of Department of Science $\&$ Technology (DST), India.\\
\end{acknowledgments}


\section*{Author Declarations}
\subsubsection*{Conflict of interest}
The authors have no conflicts to disclose.

\section*{Data Availability}
The data that support the findings of this study are available from the corresponding author upon reasonable request.

\nocite{}
\bibliography{aipsamp}

\end{document}